\documentclass[conference, letterpaper]{IEEEtran}

\newtheorem{definition}{Definition}
\newtheorem{lemma}{Lemma}
\newtheorem{theorem}{Theorem}

\usepackage[linesnumbered, ruled]{algorithm2e}

\usepackage[cmex10]{amsmath}
\usepackage{amssymb}

\usepackage{cite}

\usepackage[T1]{fontenc}

\usepackage{graphicx}

\usepackage{tikz}
\tikzset{> = latex}

\newcommand{\entropy}[1]{H ( #1 )}
\newcommand{\mean}[1]{E [ #1 ]}
\newcommand{\ocnumber}[2]{N ( #1 | #2 )} 
\newcommand{\probability}[1]{\mathrm{Pr} [ #1 ]}

\newcommand{\ccinterval}[2]{[ #1 , #2 ]} 
\newcommand{\cointerval}[2]{[ #1 , #2 )} 
\newcommand{\oointerval}[2]{( #1 , #2 )} 
\newcommand{\reals}{\mathbb{R}}
\newcommand{\integers}{\mathbb{Z}}

\newcommand{\bemap}{H_{\mathrm{b}}} 
\newcommand{\entropymap}{H}
\newcommand{\lmrate}{I_{\mathrm{LM}}}
\newcommand{\mimap}{I} 

\newcommand{\meanop}[1]{E_{#1}} 

\title
{PAC Learnability for Reliable Communication over Discrete Memoryless Channels}
\author{%
    \IEEEauthorblockN{Jiakun Liu, Wenyi Zhang}
    \IEEEauthorblockA{%
        Department of Electronic Engineering and Information Science\\
        University of Science and Technology of China, Hefei, China\\
        liujk@mail.ustc.edu.cn, wenyizha@ustc.edu.cn%
    }
    \and
    \IEEEauthorblockN{H. Vincent Poor}
    \IEEEauthorblockA{%
        Department of Electrical Engineering\\
        Princeton University, Princeton, NJ, USA\\
        poor@princeton.edu%
    }%
}

\begin{document}
    \maketitle

    \begin{abstract}
        In practical communication systems, knowledge of channel models is
        often absent, and consequently, transceivers need be designed based on
        empirical data.
        In this work, we study data-driven approaches to reliably choosing
        decoding metrics and code rates that facilitate reliable communication
        over unknown discrete memoryless channels (DMCs).
        Our analysis is inspired by the PAC (probably approximately correct)
        learning theory and does not rely on any assumptions on the statistical
        characteristics of DMCs.
        We show that a naive plug-in algorithm for choosing decoding metrics is
        likely to fail for finite training sets.
        We propose an alternative algorithm called the virtual sample algorithm
        and establish a non-asymptotic lower bound on its performance.
        The virtual sample algorithm is then used as a building block for
        constructing a learning algorithm that chooses a decoding metric and a
        code rate using which a transmitter and a receiver can reliably
        communicate at a rate arbitrarily close to the channel mutual
        information.
        Therefore, we conclude that DMCs are PAC learnable.
    \end{abstract}


\section{Introduction}
\label{s:introduction}

The application of machine learning techniques in communications has gained
broad interest in recent years
\cite{gruber2017, simeone2018, zhang2019, she2021, eldar2022}.
An advantage of machine learning techniques is that they help address channel
uncertainty.
If the statistical characteristics of a channel are known perfectly, then an
encoder and a decoder can be optimized accordingly to communicate at a rate
close to the channel capacity.
However, perfect channel knowledge is unavailable in practice.
Instead, the encoder and decoder are designed based on some empirical data
about the channel.

In this paper, we show it is possible to use empirical data to choose decoding
metrics and code rates that facilitate reliable communication over unknown
discrete memoryless channels (DMCs).
We analyze two algorithms for choosing decoding metrics: A naive plug-in
algorithm and a proposed virtual sample algorithm.
The plug-in algorithm is straightforward and intuitive, but is likely to fail
for finite training sets.
On the contrary, the virtual sample algorithm returns a desired decoding metric
with a high probability as long as the size of the training set is no less than
a certain finite value which does not depend on the channel.
Our analysis does not rely on any assumptions on the statistical
characteristics of the channel, much like that in the PAC (probably
approximately correct) learning theory \cite{valiant1984, shalevshwartz2014}.
We use the virtual sample algorithm as a building block to construct a learning
algorithm that reliably chooses a decoding metric and a code rate using which
an encoder and a decoder can communicate reliably at a rate arbitrarily close
to the channel mutual information.
This leads to a conclusion that DMCs are PAC learnable.

In the language of PAC learning, a chosen decoding metric is a hypothesis and a
risk of the hypothesis can be defined using the maximum communication rate
supported by the metric.
Some other PAC learning problems for communications have been studied in
\cite{weinberger2022, tsvieli2023, bernardo2023, tope2023}.
In \cite[Sections~II and III]{weinberger2022} and \cite{tsvieli2023}, the
hypotheses are constellations and/or their decoders, and the risks are error
probabilities or expectations of a hinge-type loss.
In \cite[Section~V]{weinberger2022}, the hypothesis is an input distribution of
a channel and the risk is the negative of the channel mutual information.
In \cite{bernardo2023}, learning-based channel codes are defined and
data-dependent achievability and converse bounds are established.
In \cite{tope2023}, input-output samples of an unknown DMC are used to
construct a compound channel that includes the DMC with a high probability.
Based on the construction, a sampling strategy is proposed to find PAC bounds
on the capacity and the corresponding input distributions.

The paper is organized as follows.
In the remaining part of Introduction, we introduce our notation and the
considered communication model.
In Section~\ref{s:formulation} we formulate the problem of how to choose
decoding metrics and code rates, and show its connection with the PAC learning
problem.
In Section~\ref{s:results} we present our theoretical results, which lead to
the PAC learnability of DMCs.
We conduct some numerical study to show the results in
Section~\ref{s:evaluations}.
Section~\ref{s:sketch} outlines the proofs of the theoretical results and
Section~\ref{s:conclusion} concludes the paper.

\subsection{Notation}
\label{u:introduction.notation}

The sets of real numbers, integers, non-negative integers and positive integers
are denoted by $\reals$, $\integers$, $\integers_{\ge 0}$ and $\integers_{> 0}$
respectively.
A finite sequence $( x_{1} , x_{2} , \cdots , x_{n} )$ is written as
$\{ x_{i} \}_{i = 1}^{n}$ or the boldface letter $\mathbf{x}$.
The number $| \{ i \in \integers | 1 \le i \le n , x_{i} = x \} |$ of the
occurrences of a symbol $x$ in $\mathbf{x}$ is written as
$\ocnumber{x}{\mathbf{x}}$.
Throughout the paper, $\mathcal{X}$ and $\mathcal{Y}$ are non-empty finite
sets.
For every probability mass function (PMF) $p$ on $\mathcal{X}$ and transition
function $w$ from $\mathcal{X}$ to $\mathcal{Y}$, $p w$ is the PMF on
$\mathcal{Y}$ defined by
$p w ( y ) = \sum_{x \in \mathcal{X}} p ( x ) w ( y | x )$, $p \times w$ is the
PMF on $\mathcal{X} \times \mathcal{Y}$ defined by
$( p \times w ) ( x , y ) = p ( x ) w ( y | x )$ and $\mimap ( p , w )$ is the
mutual information between $X$ and $Y$ when $( X , Y ) \sim p \times w$.
All entropies and mutual information are in bits.


\subsection{Communication Model}
\label{u:introduction.model}

Suppose a DMC has a transition function $w$ from $\mathcal{X}$ to
$\mathcal{Y}$.
A typical paradigm of communication over the DMC uses a codebook and a decoding
metric \cite{merhav1994, csiszar1995, salz1995, lapidoth1998, scarlett2020}.
There are $M$ codewords in the codebook and positive integers not exceeding $M$
are called messages.
If a message $m$ is to be transmitted, the encoder will input the $m$-th
codeword $\{ x_{m , t} \}_{t = 1}^{T} \in \mathcal{X}^{T}$ in the codebook into
the DMC.
The code rate of the codebook is defined as $\log_{2} ( M ) / T$.
The decoding metric is a function
$k : \mathcal{X} \times \mathcal{Y} \to \cointerval{0}{\infty}$.
Based on the sequence $\{ y_{t} \}_{t = 1}^{T} \in \mathcal{Y}^{T}$ output by
the DMC, the decoder finds the positive integer $\hat{m} \le M$ maximizing
$\prod_{t = 1}^{T} k ( x_{\hat{m} , t} , y_{t} )$ and guesses the transmitted
message as $\hat{m}$.
If $k ( x , y ) = w ( y | x )$ for all $x \in \mathcal{X}$ and
$y \in \mathcal{Y}$, the decoder is a matched decoder and $k$ is the
maximum-likelihood decoding metric.
If $k ( x , y ) \not= w ( y | x )$ for some $x \in \mathcal{X}$ and
$y \in \mathcal{Y}$, the guessed message $\hat{m}$ may be different from that
guessed by a matched decoder.
The decoder is then said to be mismatched
\cite{merhav1994, csiszar1995, salz1995, lapidoth1998, scarlett2020}.


\section{Problem Formulation}
\label{s:formulation}

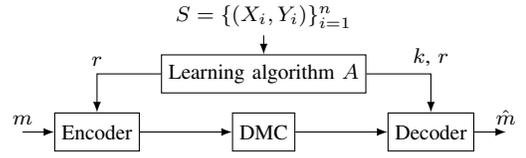
\begin{figure}[!t]
    \centering\footnotesize
    \begin{tikzpicture}[
    block/.style = {draw, minimum height = 4ex, minimum width = 1em}
]
    \matrix[row sep = 2ex, column sep = 1em]{
        &
        \node[block] (algorithm) {Learning algorithm $A$}; &
        \\
        \node[block] (encoder) {Encoder}; &
        \node[block] (channel) {DMC}; &
        \node[block] (decoder) {Decoder}; \\
    };

    \draw[<-] (encoder.west) -- ++(-1.5em, 0) node[above] {$m$};
    \draw[->] (encoder) -- (channel);
    \draw[->] (channel) -- (decoder);
    \draw[->] (decoder.east) -- ++(1.5em, 0) node[above] {$\hat{m}$};
    \draw[<-] (algorithm.north) -- ++(0, 2ex) node[above]
    {$S = \{ ( X_{i} , Y_{i} ) \}_{i = 1}^{n}$};
    \draw[->] (algorithm) -| node[above] {$r$} (encoder);
    \draw[->] (algorithm) -| node[above] {$k$, $r$} (decoder);
\end{tikzpicture}
    \caption{%
        A learning algorithm accepts $n$ i.i.d. input-output pairs of the DMC
        as input and returns a decoding metric $k$ and a code rate $r$, which
        are used by an encoder and a decoder.}
    \label{f:formulation.coding}
\end{figure}

We study how to use empirical data to choose a decoding metric and a code rate
for a DMC with a transition function $w$ from $\mathcal{X}$ to $\mathcal{Y}$
and an input distribution $p$ on $\mathcal{X}$.
We assume that $\mathcal{X}$ and $\mathcal{Y}$ are known, but there is no
knowledge about $w$ or $p$.
Furthermore, we do not assume parametric forms for $w$ or $p$.
We have a training set $S$ of independent and identically distributed (i.i.d.)
input-output pairs $( X_{1} , Y_{1} )$, $( X_{2} , Y_{2} )$, $\cdots$,
$( X_{n} , Y_{n} )$ of the DMC.
The training set is input into a learning algorithm $A$, which is a mapping
from the training set to $( k , r )$, as shown in
\figurename~\ref{f:formulation.coding}.
The communication will be at the rate $r$ and the decoder will use the decoding
metric $k$.

If the size $n$ of the training set is sufficiently large, then we will have
precise knowledge of $w$, and can let $k$ be the maximum-likelihood decoding
metric.
Then under the input distribution $p$, the encoder and decoder can use a proper
codebook and the decoding metric to communicate at the highest rate
$\mimap ( p , w )$.
However, if $n$ is not large, it is impossible and unnecessary to know $w$
precisely and $k$ is good enough as long as it facilitates high-rate
communication.
Hence we do not evaluate $k$ using its closeness to $w$.
Instead, we evaluate $k$ using the highest rate achievable by $k$ and random
codebooks with the constant composition $p$, i.e. define LM rate
\cite{scarlett2020}
\begin{equation}
    \lmrate ( p , w , k )
    = \min_{\tilde{w} \in \tilde{W} ( p , w )} \mimap ( p , \tilde{w} ) ,
    \label{e:formulation.lmrate}
\end{equation}
where $\tilde{W} ( p , w )$ is the set of transition functions $\tilde{w}$ from
$\mathcal{X}$ to $\mathcal{Y}$ satisfying $p \tilde{w} = p w$ and
\begin{equation}
    \meanop{( X , Y ) \sim p \times \tilde{w}} \log_{2} ( k ( X , Y ) )
    \ge \meanop{( X , Y ) \sim p \times w} \log_{2} ( k ( X , Y ) ) .
    \label{e:formulation.means}
\end{equation}
As the length of the codewords in the random codebook tends to infinity, the
error probability tends to one at rates higher than $\lmrate ( p , w , k )$ and
tends to zero at rates lower than $\lmrate ( p , w , k )$
\cite{csiszar1981, hui1983, merhav1994, csiszar1995, scarlett2020}.
In general $0 \le \lmrate ( p , w , k ) \le \mimap ( p , w )$.
If $k$ is the maximum-likelihood decoding metric, then
$\lmrate ( p , w , k ) = \mimap ( p , w )$.
We can see $\lmrate ( p , w , k )$ is a lower bound on the highest rate
achievable by decoders with the metric $k$.
The LM rate is generally not the highest rate achieved by a mismatched decoder
(see, e.g., \cite{lapidoth1996}), and we use the LM rate in our study merely as
a tool for establishing the PAC learnability result.

\paragraph*{The Problem and Its Connection With PAC Learning}
In the language of PAC learning, $k$ is a hypothesis and
$- \lmrate ( p , w , k )$ is the risk of $k$.
Our problem is similar to a PAC learning problem except that our learning
algorithm should return a code rate in addition to the hypothesis $k$.
The code rate $r$ should be less than or equal to $\lmrate ( p , w , k )$ to
ensure asymptotically reliable decoding.
On the other hand, $r$ is desired to be as high as possible.
Therefore, $- r$ should be a good estimate and a tight upper bound of the risk.
Also note that when $k$ is the maximum-likelihood decoding metric, the LM rate
is equal to $\mimap ( p , w )$, or equivalently, the lowest risk is
$- \mimap ( p , w )$.
In summary, we hope to find a learning algorithm that returns a decoding metric
$k$ and a code rate $r$ satisfying
$\mimap ( p , w ) - \epsilon \le r \le \lmrate ( p , w , k )$ with a high
probability $1 - \delta$, where $\epsilon$ and $\delta$ are small positive
numbers.

In our problem formulation, the learning algorithm can be implemented at the
decoder, and the encoder only needs to be able to adjust the code rate.
No interactions between the encoder and decoder are needed except that the
decoder needs to inform the encoder the learned rate $r$ (see
\figurename~\ref{f:formulation.coding}).

We also remark that universal decoding \cite{ziv1985, csiszar2011} may also be
viewed as a data-driven approach, because the decoder does not need to know the
channel.
Although universal decoding can achieve the channel capacity and even the
error exponent of a channel, the implementation of universal decoding is
usually believed to be of high complexity.
In contrast, the decoder structure considered in our paradigm is compatible
with existing channel coding techniques used in practical systems.
Furthermore, in our problem formulation, the code rate is the target to be
learned, rather than prescribed a priori.

\section{Theoretical Results}
\label{s:results}

Let $s \in ( \mathcal{X} \times \mathcal{Y} )^{n}$ be a training set.
There exist a PMF $p_{s}$ on $\mathcal{X}$ and a transition function $w_{s}$
from $\mathcal{X}$ to $\mathcal{Y}$ satisfying
$\ocnumber{x , y}{s} = n p_{s} ( x ) w_{s} ( y | x )$ for all
$x \in \mathcal{X}$ and $y \in \mathcal{Y}$.
We can see $p_{s}$ and $w_{s}$ are empirical distributions of $s$.
Denote $k_{s} ( x , y ) = w_{s} ( y | x )$.
Since $
    \lmrate ( p_{s} , w_{s} , k )
    \le \mimap ( p_{s} , w_{s} )
    = \lmrate ( p_{s} , w_{s} , k_{s} )
$ for all $k : \mathcal{X} \times \mathcal{Y} \to \cointerval{0}{\infty}$,
$k = k_{s}$ minimizes the empirical risk $- \lmrate ( p_{s} , w_{s} , k )$.
It appears to be reasonable to choose $k_{s}$ as the decoding metric, as done
by the plug-in algorithm.

\begin{definition}
    \label{def:results.pia} 
    A learning algorithm $A$ is a plug-in algorithm if for every
    $n \in \integers_{> 0}$ and $s \in ( \mathcal{X} \times \mathcal{Y} )^{n}$
    as the input, $A$ returns a decoding metric
    $k : \mathcal{X} \times \mathcal{Y} \to \cointerval{0}{\infty}$ such that
    there exists a PMF $p_{s}$ on $\mathcal{X}$ satisfying
    $\ocnumber{x , y}{s} = n p_{s} ( x ) k ( x , y )$ for all
    $x \in \mathcal{X}$ and $y \in \mathcal{Y}$.
\end{definition}

All plug-in algorithms are empirical risk minimization algorithms.
The following theorem is about the performance of plug-in algorithms.

\begin{theorem}
    \label{thm:results.pia}
    If $| \mathcal{X} | \ge 2$, $| \mathcal{Y} | \ge 2$, $0 < \epsilon < 1$,
    and $0 < \delta < 1 / 2$, then for every $n \in \integers_{> 0}$, there
    exist a PMF $p$ on $\mathcal{X}$ and a transition function $w$ from
    $\mathcal{X}$ to $\mathcal{Y}$ satisfying
    \begin{enumerate}
        \item $\mimap ( p , w ) > \epsilon$;
        \item If $S$ is a training set of $n$ i.i.d. pairs generated by
        $p \times w$ and a plug-in algorithm returns $K$ for the input $S$,
        then
        \begin{equation}
            \probability{\lmrate ( p , w , K ) = 0} > \delta .
            \label{e:results.p0lmr} 
        \end{equation}
    \end{enumerate}
\end{theorem}

We write the training set and decoding metric in Theorem~\ref{thm:results.pia}
as the upper-case letters $S$ and $K$ to indicate they are random.
For fixed $p$ and $w$, by the law of large numbers, $K$ converges to the
maximum-likelihood decoding metric as $n \to \infty$, implying
$\lmrate ( p , w , K ) \to \mimap ( p , w )$ as $n \to \infty$.
On the other hand, Theorem~\ref{thm:results.pia} says that for every fixed $n$
there exist $p$ and $w$ such that the plug-in algorithm fails with a high
probability.
In this sense, the plug-in algorithm does not address channel uncertainty.

A problem of the plug-in algorithm is that if there is no $( x , y )$ in the
training set, then the algorithm is likely to return a decoding metric $k$
satisfying $k ( x , y ) = 0$.
If the probability $p ( x ) w ( y | x )$ is in fact positive, then
$\lmrate ( p , w , k ) = 0$ (See Lemma~\ref{lem:sketch.c0lmr} in
Section~\ref{s:sketch}).
In fact, we prove Theorem~\ref{thm:results.pia} using this idea.
The problem can be avoided by adding some virtual instances of $( x , y )$ to
the training set, as done by Laplace's law \cite{krichevskiy1998}.
This inspires the virtual sample algorithm.

\begin{definition}
    \label{def:results.vsa} 
    For $\alpha \in \reals$, the $\alpha$-virtual sample algorithm is the
    learning algorithm that, for every $n \in \integers_{> 0}$ and
    $s \in ( \mathcal{X} \times \mathcal{Y} )^{n}$ as the input, returns a
    decoding metric
    $k : \mathcal{X} \times \mathcal{Y} \to \cointerval{0}{\infty}$ satisfying
    $k ( x , y ) = \ocnumber{x , y}{s} + n^{\alpha}$ for all
    $x \in \mathcal{X}$ and $y \in \mathcal{Y}$.
\end{definition}

The decoding metric $k$ in Definition~\ref{def:results.vsa} can be viewed
as a smoothed estimate of $p \times w$.
The hyper-parameter $\alpha$ controls the extent of smoothing.

\begin{theorem}
    \label{thm:results.vsa}
    If $1 / 2 < \alpha < 1$, $\epsilon > 0$, $0 < \delta < 1$, $p$ is a PMF on
    $\mathcal{X}$, $w$ is a transition function from $\mathcal{X}$ to
    $\mathcal{Y}$, $S$ is a training set of $n$ i.i.d. pairs generated by
    $p \times w$,
    \begin{align}
        n
        & \ge \left(
            \frac{| \mathcal{X} | | \mathcal{Y} |}{\epsilon \ln ( 2 )}
        \right)^{1 / ( 1 - \alpha )} , \label{e:results.scac} \\
        n
        & \ge \left( \frac{1}{2} \ln \left(
            \frac{| \mathcal{X} | | \mathcal{Y} |}{\delta}
        \right) \right)^{1 / ( 2 \alpha - 1 )} \label{e:results.scpc}
    \end{align}
    and the $\alpha$-virtual sample algorithm returns $K$ for the input $S$,
    then
    \begin{equation}
        \probability{\mimap ( p , w ) - \epsilon < \lmrate ( p , w , K )}
        > 1 - \delta .
        \label{e:results.phlmr} 
    \end{equation}
\end{theorem}

Theorem~\ref{thm:results.vsa} provides a performance guarantee for the
$\alpha$-virtual sample algorithm that does not rely on the specific transition
function $w$ or the input distribution $p$ of the DMC.
In this sense, the algorithm addresses channel uncertainty.
We can also say that the hypothesis class of all functions from
$\mathcal{X} \times \mathcal{Y}$ to $\cointerval{0}{\infty}$ is PAC learnable,
and that the virtual sample algorithm is a successful PAC learner for the
hypothesis class.

The difference between the performance of the algorithms can be interpreted
using the philosophy that the unseen (i.e. $\ocnumber{x , y}{S} = 0$ here) is
not equivalent to the impossible (i.e. $p ( x ) w ( y | x ) = 0$ here), which
has been extensively studied in a series of works like \cite{orlitsky2003} and
\cite{valiant2017}.

Once a decoding metric is returned by the virtual sample algorithm, we can
choose a code rate using an estimate of the mutual information
$\mimap ( p , w )$.
We estimate $\mimap ( p , w )$ using the Miller-Madow corrected entropy
estimator \cite{miller1955}.
Given a realization $\mathbf{x}$ and the alphabet size $l$ of a sequence of $n$
i.i.d. discrete random variables, the Miller-Madow corrected entropy estimator
$\hat{H}_{\mathrm{MM}}$ estimates the entropy of each variable as
\begin{align}
    \hat{H}_{\mathrm{MM}} ( \mathbf{x} , l )
    ={} & - \sum_{x , \ocnumber{x}{\mathbf{x}} > 0}
    \frac{\ocnumber{x}{\mathbf{x}}}{n}
    \log_{2} \left( \frac{\ocnumber{x}{\mathbf{x}}}{n} \right) \notag \\
    & + \frac{l - 1}{2 n} \log_{2} ( e ) .
\end{align}
The $( ( l - 1 ) / 2 n ) \log_{2} ( e )$ correction term reduces the asymptotic
bias rate of the estimate to $o ( 1 / n )$.
A series of more advanced entropy estimators
\cite{paninski2003, valiant2017, jiao2015, wu2016, verdu2019} has been
proposed, but they are designed for asymptotically large alphabets or have
extra hyper-parameters.
It is easier to implement the Miller-Madow corrected estimator and to analyze
the performance of it for fixed small alphabets.
Hence we adopt the Miller-Madow corrected estimator for technical convenience.
We call the algorithm choosing code rates using the estimate of
$\mimap ( p , w )$ the virtual sample and entropy estimation (VSEE) algorithm.

\begin{definition}
    \label{def:results.vseea}
    For $\alpha \in \reals$, $\beta \in \reals$, $n \in \integers_{> 0}$ and $
        s
        = \{ ( x_{i} , y_{i} ) \}_{i = 1}^{n}
        \in ( \mathcal{X} \times \mathcal{Y} )^{n}
    $ as the input, the $( \alpha , \beta )$-VSEE algorithm
    \begin{enumerate}
        \item Gets a decoding metric $k$ by calling the $\alpha$-virtual sample
        algorithm with the input $s$;
        \item Computes the estimate $
            \hat{I}
            = \hat{H}_{\mathrm{MM}} (
                \{ x_{i} \}_{i = 1}^{n} ,
                | \mathcal{X} |
            )
            + \hat{H}_{\mathrm{MM}} (
                \{ y_{i} \}_{i = 1}^{n} ,
                | \mathcal{Y} |
            )
            - \hat{H}_{\mathrm{MM}} ( s , | \mathcal{X} | | \mathcal{Y} | )
        $ of mutual information;
        \item Returns $( k , \hat{I} - n^{- \beta} )$.
    \end{enumerate}
\end{definition}

\begin{theorem}
    \label{thm:results.vseea}
    There exists a function $
        \nu :
        \oointerval{1 / 2}{\infty}
        \times \oointerval{0}{\infty}
        \times \oointerval{0}{\infty}
        \times \oointerval{0}{1}
        \to \integers_{\ge 0}
    $ such that
    \begin{equation}
        \probability{
            \mimap ( p , w ) - \epsilon \le R \le \lmrate ( p , w , K )
        }
        > 1 - \delta
    \end{equation}
    if $\alpha > 1 / 2$, $\beta > 0$, $\alpha + \beta < 1$, $\epsilon > 0$,
    $0 < \delta < 1$, $p$ is a PMF on $\mathcal{X}$, $w$ is a transition
    function from $\mathcal{X}$ to $\mathcal{Y}$, $S$ is a training set of
    $n \ge \nu ( \alpha , \beta , \epsilon , \delta )$ i.i.d. pairs generated
    by $p \times w$ and the $( \alpha , \beta )$-VSEE algorithm returns
    $( K , R )$ for the input $S$.
\end{theorem}

The encoder and decoder in Section~\ref{s:formulation} can communicate reliably
at a high rate whenever
$\mimap ( p , w ) - \epsilon \le R \le \lmrate ( p , w , K )$.
Since this is all we want to achieve by learning, we can conclude that DMCs
with the alphabets $\mathcal{X}$ and $\mathcal{Y}$ are PAC learnable, though
the DMCs do not compose a hypothesis class.
The existence of such a $\nu$ suffices to establish the PAC learnability.
The explicit characterization of $\nu$ is left for future research.

    \section{Empirical Evaluations of Algorithms}
    \label{s:evaluations}
    \subsection{Virtual Sample Algorithm}
\label{u:evaluations.vsa}

Theorem~\ref{thm:results.vsa} provides a rule of thumb for choosing the
hyper-parameter of the virtual sample algorithm.
Suppose we have fixed $\epsilon \in \oointerval{0}{\infty}$ and
$\delta \in \oointerval{0}{1}$.
We hope to choose a decoding metric for a DMC with an unknown transition
function $w$ and an unknown input distribution $p$.
Theorem~\ref{thm:results.vsa} says that the $\alpha$-virtual sample
algorithm returns a decoding metric $K$ satisfying
$\mimap ( p , w ) - \epsilon < \lmrate ( p , w , K )$ with a probability
greater than $1 - \delta$ if $1 / 2 < \alpha < 1$ and there are at least $
    \nu_{\alpha}
    = \max ( e^{\zeta / ( 1 - \alpha )} , e^{\eta / ( 2 \alpha - 1 )} )$
instances in the training set, where
\begin{equation}
    \zeta
    = \ln \left(
        \frac{| \mathcal{X} | | \mathcal{Y} |}{\epsilon \ln ( 2 )}
    \right)
\end{equation}
and
\begin{equation}
    \eta
    = \ln \left(
        \frac{1}{2}
        \ln \left( \frac{| \mathcal{X} | | \mathcal{Y} |}{\delta} \right)
    \right) .
\end{equation}
Of course we should choose an $\alpha$ that minimizes $\nu_{\alpha}$.
Since $\epsilon$ and $\delta$ are typically small, $\zeta$ and $\eta$ are
typically positive.
Then $e^{\zeta / ( 1 - \alpha )}$ increases and $e^{\eta / ( 2 \alpha - 1 )}$
decreases with $\alpha \in \oointerval{1 / 2}{1}$.
The $\alpha$ minimizing $\nu_{\alpha}$ satisfies
$e^{\zeta / ( 1 - \alpha )} = e^{\eta / ( 2 \alpha - 1 )}$ and consequently
$\alpha = ( \zeta + \eta ) / ( 2 \zeta + \eta )$.

Let us consider the following example: $\mathcal{X} = \{ 0 , 1 \}$,
$\mathcal{Y} = \{ 0 , 1 , 2 \}$, $\epsilon = 0.05$ and $\delta = 0.1$.
When $\alpha \approx 0.5325$, $\nu_{\alpha}$ is minimized and approximately
$6.136 \times 10^{4}$.
A training set of size $6.136 \times 10^{4}$ seems too large for DMCs with such
small alphabets.
However, note that this size provides a worst-case guarantee on the performance
of the 0.5325-virtual sample algorithm for all possible $p$ and $w$.
Our experiment shows that for some common $p$ and $w$, the 0.5325-virtual
sample algorithm performs well even if the training set is much smaller.

\begin{figure}[!t]
    \centering
    \includegraphics[scale = 0.35]{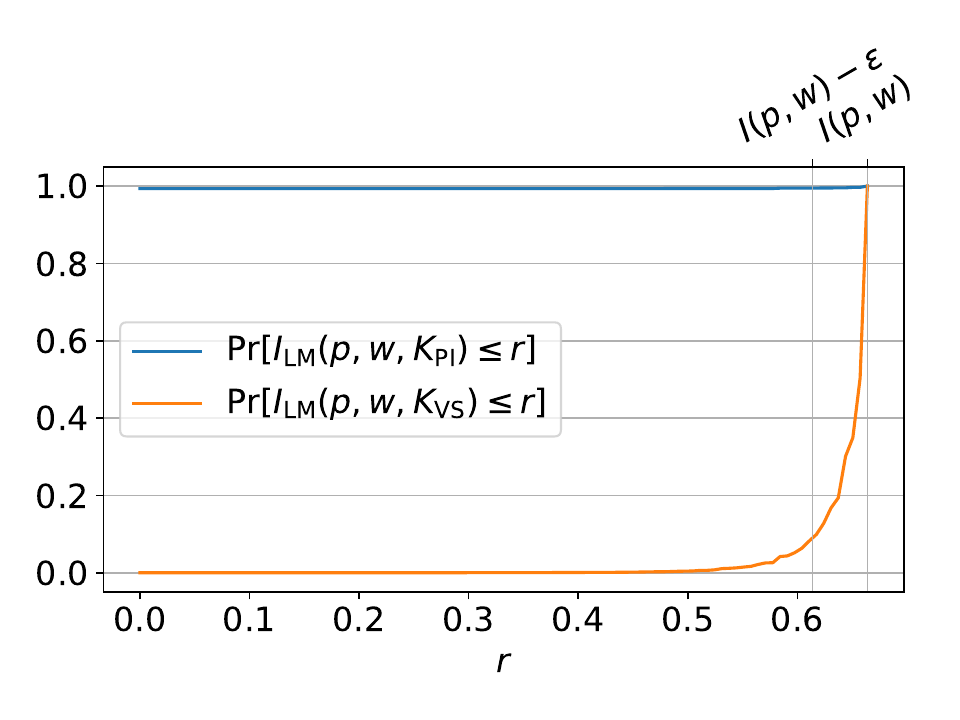}
    \caption{%
        The CDFs of $\lmrate ( p , w , K_{\mathrm{PI}} )$ and
        $\lmrate ( p ,w , K_{\mathrm{VS}} )$.}
    \label{f:evaluations.vsa.comparison}
\end{figure}

Let $p ( 0 ) = p ( 1 ) = 1 / 2$,
\begin{equation}
    \begin{bmatrix}
        w ( 0 | 0 ) & w ( 1 | 0 ) & w ( 2 | 0 ) \\
        w ( 0 | 1 ) & w ( 1 | 1 ) & w ( 2 | 1 )
    \end{bmatrix}
    = \begin{bmatrix}
        0.86 & 0.1 & 0.04 \\
        0.04 & 0.1 & 0.86
    \end{bmatrix} ,
    \label{e:evaluations.vsa.transfunc} 
\end{equation}
$S = \{ ( X_{i} , Y_{i} ) \}_{i = 1}^{12}$ be a sequence of i.i.d. pairs
generated by $p \times w$, $\mathbf{X} = \{ X_{i} \}_{i = 1}^{12}$ and
\begin{equation}
    K_{\mathrm{PI}} ( x , y )
    = \begin{cases}
        \ocnumber{x , y}{S} / \ocnumber{x}{\mathbf{X}} , &
        \ocnumber{x}{\mathbf{X}} > 0 , \\
        1 / 3 , &
        \ocnumber{x}{\mathbf{X}} = 0
    \end{cases} \label{e:evaluations.vsa.pia}
\end{equation}
for all $x \in \mathcal{X}$ and $y \in \mathcal{Y}$.
We can see \eqref{e:evaluations.vsa.pia} defines a plug-in algorithm.
Suppose the 0.5325-virtual sample algorithm returns $K_{\mathrm{VS}}$ for the
input $S$.
We compute the cumulative distribution functions (CDFs) of
$\lmrate ( p , w , K_{\mathrm{PI}} )$ and $\lmrate ( p ,w , K_{\mathrm{VS}} )$
using the techniques described in Appendix~\ref{s:techniques} and show them in
\figurename~\ref{f:evaluations.vsa.comparison}.
According to \figurename~\ref{f:evaluations.vsa.comparison}, the probability of
$\mimap ( p , w ) - \epsilon < \lmrate ( p , w , K_{\mathrm{VS}} )$ exceeds
$1 - \delta = 0.9$.
Hence a training set of size 12 is sufficient for the 0.5325-virtual sample
algorithm to choose a good decoding metric for the DMC.
On the contrary, the probability of $\lmrate ( p , w , K_{\mathrm{PI}} ) = 0$
is close to 1, confirming what we have predicted using
Theorem~\ref{thm:results.pia}.

\begin{figure}[!t]
    \centering
    \includegraphics[scale = 0.35]
    {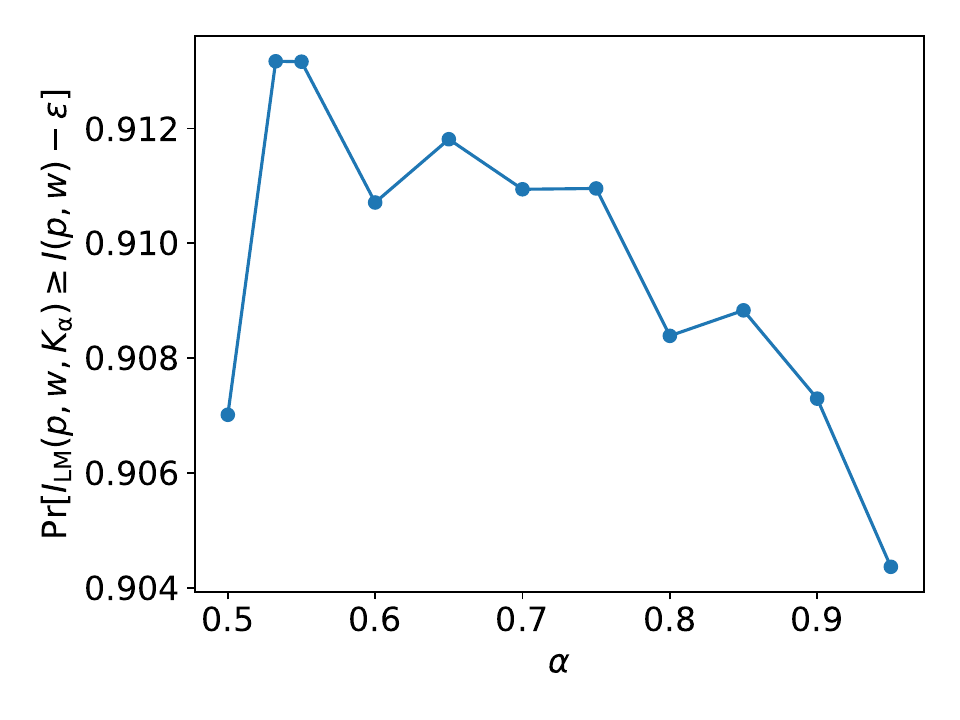}
    \caption{%
        The probabilities of \eqref{e:evaluations.vsa.success} for
        $\alpha \in \{ 0.5 , 0.5325 , 0.55 , 0.6 , \cdots , 0.95 \}$.}
    \label{f:evaluations.vsa.probabilities}
\end{figure}

To verify our choice of $\alpha$, we repeat the experiment on the
$\alpha$-virtual sample algorithm for every
$\alpha \in A = \{ 0.5 , 0.55 , \cdots , 0.95 \}$.
For $\alpha \in A \cup \{ 0.5325 \}$, suppose the $\alpha$-virtual sample
algorithm returns $K_{\alpha}$ for the input $S$.
\figurename~\ref{f:evaluations.vsa.probabilities} shows the probabilities of
\begin{equation}
    \mimap ( p , w ) - \epsilon \le \lmrate ( p , w , K_{\alpha} ) .
    \label{e:evaluations.vsa.success}
\end{equation}
Although the specific values of $\alpha \in A \cup \{ 0.5325 \}$ do not have a
large impact on the probability, $\alpha = 0.5325$ maximizes the probability of
\eqref{e:evaluations.vsa.success}.
Therefore, Theorem~\ref{thm:results.vsa} provides a good rule of thumb for
choosing $\alpha$.


\subsection{VSEE Algorithm}
\label{u:evaluations.vseea}

\begin{figure}[!t]
    \centering
    \includegraphics[scale = 0.35]{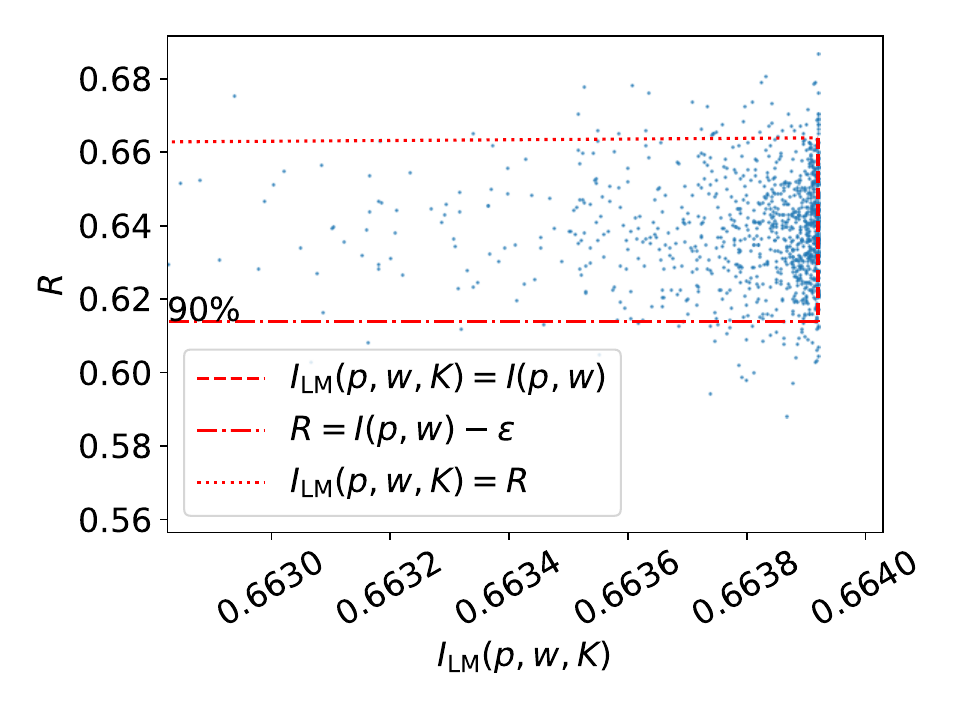}
    \caption{%
        The realizations of $( \lmrate ( p , w , K ) , R )$ in 1000 independent
        experiments.
        The relation \eqref{e:evaluations.vseea.success} is satisfied if and
        only if the blue dot representing the realization lies in the region
        surrounded by the red lines.}
    \label{f:evaluations.vseea.realizations}
\end{figure}

Consider the same $\mathcal{X}$, $\mathcal{Y}$, $\epsilon$, $\delta$, $p$ and
$w$ as those in Section~\ref{u:evaluations.vsa}.
Since the 0.5325-virtual sample algorithm performs well, we let the first
hyper-parameter $\alpha$ of the VSEE algorithm be 0.5325.
Let $S = \{ ( X_{i} , Y_{i} ) \}_{i = 1}^{3500}$ be a sequence of i.i.d. pairs
generated by $p \times w$.
We run the (0.5325, 0.45)-VSEE algorithm with the input $S$ to obtain
$( K , R )$ in each of 1000 independent experiments and show the realizations
of $( \lmrate ( p , w , K ) , R )$ in
\figurename~\ref{f:evaluations.vseea.realizations}.

The relation
\begin{equation}
    \mimap ( p , w ) - \epsilon \le R \le \lmrate ( p , w , K )
    \label{e:evaluations.vseea.success}
\end{equation}
is satisfied in approximately 90\% (i.e. $1 - \delta$) of the realizations.
The size 3500 is much smaller than the size
$\nu_{\alpha} \approx 6.136 \times 10^{4}$ predicted in
Section~\ref{u:evaluations.vsa}, but still seems too large for DMCs with the
small alphabets $\mathcal{X}$ and $\mathcal{Y}$.
\figurename~\ref{f:evaluations.vseea.realizations} also shows that the
realizations of $\lmrate ( p , w , K )$ are very close to $\mimap ( p , w )$
and that the realizations of $R$ spread the interval $\ccinterval{0.59}{0.68}$.
Hence the performance of the VSEE algorithm is mainly affected by the error of
the estimate of $\mimap ( p , w )$, confirming the fact that it is difficult to
estimate entropy using empirical data
\cite{paninski2003, verdu2019, valiant2017}.
We can also see any good estimate of $\mimap ( p , w )$ is also a good estimate
of $\lmrate ( p , w , K )$.
Hence it may be more difficult to estimate LM rates than to choose decoding
metrics with high LM rates using empirical data.


\section{Proof Sketch of the Theorems}
\label{s:sketch}

We use the following lemmas in the proof of Theorem~\ref{thm:results.pia}.

\begin{lemma}
    \label{lem:sketch.c0lmr} 
    If $p$ is a PMF on $\mathcal{X}$, $w$ is a transition function from
    $\mathcal{X}$ to $\mathcal{Y}$, $k$ is a function from
    $\mathcal{X} \times \mathcal{Y}$ to $\cointerval{0}{\infty}$,
    $a \in \mathcal{X}$, $b \in \mathcal{Y}$, $p ( a ) w ( b | a ) > 0$ and
    $k ( a , b ) = 0$, then $\lmrate ( p , w , k ) = 0$.
\end{lemma}

\begin{lemma}
    \label{lem:sketch.p0m} 
    If $p$ is a PMF on $\mathcal{X}$, $w$ is a transition function from
    $\mathcal{X}$ to $\mathcal{Y}$, $a \in \mathcal{X}$, $b \in \mathcal{Y}$,
    $S$ is a training set of $n$ i.i.d. pairs generated by $p \times w$ and a
    plug-in algorithm returns $K$ for the input $S$, then
    \begin{equation}
        \probability{K ( a , b ) = 0}
        \ge ( 1 - p ( a ) w ( b | a ) )^{n} - ( 1 - p ( a ) )^{n} .
        \label{e:sketch.p0m}
    \end{equation}
\end{lemma}

Given $\epsilon \in \oointerval{0}{1}$, $\delta \in \oointerval{0}{1 / 2}$ and
$n \in \integers_{> 0}$, we construct the PMF $p$ and transition function $w$
in Theorem~\ref{thm:results.pia} as follows.
For any PMF $p$ on $\mathcal{X}$ and transition function $w$ from $\mathcal{X}$
to $\mathcal{Y}$ satisfying $p ( a ) w ( b | a ) > 0$, by
Lemmas~\ref{lem:sketch.c0lmr} and \ref{lem:sketch.p0m},
$( 1 - p ( a ) w ( b | a ) )^{n} - ( 1 - p ( a ) )^{n}$ is a lower bound of
$\probability{\lmrate ( p , w , K ) = 0}$.
The probability can be large enough as long as $w ( b | a )$ is small enough.
We choose such a $w$ and let $p$ be a PMF on $\mathcal{X}$ with only two
support points, each of which has probability $1 / 2$.

Lemma~\ref{lem:sketch.c0lmr} is proved using \eqref{e:formulation.lmrate}.
Alternatively, we can motivate Lemma~\ref{lem:sketch.c0lmr} by showing that any
random codebook with the constant composition $p$ and a positive rate leads to
a non-negligible probability of error.
If the DMC outputs $\{ y_{t} \}_{t = 1}^{T}$ for a codeword
$\{ x_{m , t} \}_{t = 1}^{T}$ and there is at least one positive integer
$t \le T$ satisfying $x_{m , t} = a$ and $y_{t} = b$, then because
$k ( a , b ) = 0$, $\prod_{t = 1}^{T} k ( x_{m , t} , y_{t} ) = 0$.
Hence the decoder with the metric $k$ does not guess the transmitted message to
be $m$, causing an error.
Since $p ( a ) w ( b | a ) > 0$, the probability of the error is not
negligible.

We use the following lemma in the proof of Theorem~\ref{thm:results.vsa}.

\begin{lemma}
    \label{lem:sketch.chlmr} 
    If $p$ is a PMF on $\mathcal{X}$, $w$ is a transition function from
    $\mathcal{X}$ to $\mathcal{Y}$, $k$ is a function from
    $\mathcal{X} \times \mathcal{Y}$ to $\cointerval{0}{\infty}$,
    $\epsilon > 0$ and
    \begin{equation}
        2^{\epsilon / | \mathcal{Y} | p w ( y )} p w ( y ) k ( x , y )
        > p ( x ) w ( y | x ) \sum_{x' \in \mathcal{X}} k ( x' , y )
        \label{e:sketch.chlmr}
    \end{equation}
    for all $x \in \mathcal{X}$ and $y \in \mathcal{Y}$ satisfying
    $p ( x ) w ( y | x ) > 0$, then
    $\mimap ( p , w ) - \epsilon < \lmrate ( p , w , k )$.
\end{lemma}

Lemma~\ref{lem:sketch.chlmr} is a consequence of the dual expression of
$\lmrate ( p , w , k )$ \cite{merhav1994, scarlett2020}:
\begin{equation}
    \sup_{\substack{
        \theta \in \cointerval{0}{\infty} \\
        a : \mathcal{X} \to \reals
    }}
    \meanop{( X , Y ) \sim p \times w}
    \log_{2} \bigg( \frac{
        k^{\theta} ( X , Y ) 2^{a ( X )}
    }{
        \sum\limits_{x \in \mathcal{X}}
        p ( x )
        k^{\theta} ( x , Y )
        2^{a ( x )}
    } \bigg) .
\end{equation}
To prove Theorem~\ref{thm:results.vsa}, we use the Hoeffding inequality
\cite{hoeffding1963} to show that \eqref{e:sketch.chlmr} holds with a high
probability, where $k$ is the decoding metric returned by the virtual sample
algorithm.
The dependence of the probability on $p$ and $w$ is eliminated by some delicate
uses of the fact that $e^{u} - 1 > u$ for all $u > 0$.

The proof of Theorem~\ref{thm:results.vseea} is an analysis of the VSEE
algorithm.
From Theorem~\ref{thm:results.vsa} and Section~\ref{s:evaluations} we can see
any good estimate of $\mimap ( p , w )$ is also a good estimate of
$\lmrate ( p , w , K )$, where $K$ is the decoding metric returned by the
virtual sample algorithm.
However, the estimate tends to be higher than $\lmrate ( p , w , K )$, while we
need an estimate that is lower than $\lmrate ( p , w , K )$ with a high
probability, as required by the problem formulation (see
Section~\ref{s:formulation}).
This problem is solved in the VSEE algorithm by choosing the code rate as an
estimate of $\mimap ( p , w )$ minus $n^{- \beta}$ for a proper $\beta$.
We use existing results on entropy estimation \cite{antos2001, paninski2003} to
show that the estimate of $\mimap ( p , w )$ can be accurate enough.
Hence, the code rate is high enough and lower than $\lmrate ( p , w , K )$ with
a high probability.

The lemmas and theorems are proved in Appendices~\ref{s:lemmaproofs} and
\ref{s:theoremproofs} respectively.

    \section{Conclusion}
\label{s:conclusion}

We have shown that it is possible to use empirical data to choose decoding
metrics and code rates that facilitate reliable communication over unknown
DMCs.
This leads to the conclusion that DMCs are PAC learnable.
The virtual sample algorithm for choosing decoding metrics works well for
training sets with practical sizes.
The design of algorithms that choose desirable code rates using small training
sets is left for future research.
Another important problem, also left for future research, is learning an almost
capacity-achieving input probability distribution, under a given training set
with a limited amount of interaction between encoder and decoder.



\section*{Acknowledgement}

The work of J. Liu and W. Zhang was supported in part by the National Natural
Science Foundation of China under Grant 62231022, and the work of H. V. Poor
was supported in part by the U.S National Science Foundation under Grants
CNS-2128448 and ECCS-2335876.

    \bibliography{IEEEabrv, bibliography}

    \clearpage
    \appendices

    \section{Lemmas Used in Proofs of the Theorems in Section~\ref{s:results}}
\label{s:lemmaproofs}

In addition to the lemmas in Section~\ref{s:sketch}, we need the following
lemma to prove Theorem~\ref{thm:results.vseea}.

\begin{lemma}
    \label{lem:lemmaproofs.eeerror} 
    If i.i.d. random variables $Z_{1}$, $Z_{2}$, $\cdots$, $Z_{n}$ take values
    on a finite set $\mathcal{Z}$, $n \ge 2$ and $0 < \delta < 1$, then with a
    probability no less than $1 - \delta$,
    \begin{align}
        & |
            \hat{H}_{\mathrm{MM}} ( \{ Z_{i} \}_{i = 1}^{n} , | \mathcal{Z} | )
            - \entropy{Z_{1}}
        | \notag \\
        & \le \frac{| \mathcal{Z} | - 1}{2 n}
        \log_{2} ( e )
        + \log_{2} ( n )
        \sqrt{\frac{2}{n} \ln \left( \frac{2}{\delta} \right)} .
        \label{e:lemmaproofs.eeerror}
    \end{align}
\end{lemma}

We prove the lemmas in the following.

\subsection{Proof of Lemma~\ref{lem:sketch.c0lmr}}
\label{u:lemmaproofs.c0lmr}

Define a transition function $\tilde{w}$ from $\mathcal{X}$ to $\mathcal{Y}$
such that $\tilde{w} ( y | x ) = p w ( y )$ for all $x \in \mathcal{X}$ and
$y \in \mathcal{Y}$.
Then we have $p \tilde{w} = p w$.
By the assumptions $p ( a ) w ( b | a ) > 0$ and $k ( a , b ) = 0$, we have
\begin{equation}
    \meanop{( X , Y ) \sim p \times w} \log_{2} ( k ( X , Y ) ) = - \infty ,
\end{equation}
which implies \eqref{e:formulation.means}.
Hence, $\lmrate ( p , w , k ) \le \mimap ( p , \tilde{w} )$.
Since $\tilde{w} ( \cdot | x )$ does not depend on $x$, random variables with
the joint distribution $p \times \tilde{w}$ are independent.
Hence, we have $\mimap ( p , \tilde{w} ) = 0$ and consequently
$\lmrate ( p , w , k ) = 0$.

\subsection{Proof of Lemma~\ref{lem:sketch.p0m}}
\label{u:lemmaproofs.p0m}

Let $S = \{ ( X_{i} , Y_{i} ) \}_{i = 1}^{n}$.
Define
\begin{equation}
    \mathcal{Z}
    = \{ ( x , y ) \in \mathcal{X} \times \mathcal{Y} | x = a , y \not= b \} ,
\end{equation}
$\mathcal{I} = \{ 1 , 2 , \cdots , n \}$ and
$L = \{ i \in \mathcal{I} | X_{i} = a \}$.
We can see $K ( a , b ) = 0$ if $L \not= \emptyset$ and $Y_{i} \not= b$ for any
$i \in L$.
Hence
\begin{equation}
    \probability{K ( a , b ) = 0}
    \ge \sum_{\Lambda \subseteq \mathcal{I} , \Lambda \not= \emptyset}
    \probability{A_{\Lambda}} ,
    \label{e:lemmaproofs.p0m.psp} 
\end{equation}
where $A_{\Lambda}$ denotes the event that $L = \Lambda$ and $Y_{i} \not= b$
for any $i \in L$.
For $\Lambda \subseteq \mathcal{I}$, $A_{\Lambda}$ equals the event that
$( X_{i} , Y_{i} ) \in \mathcal{Z}$ for all $i \in \Lambda$ and $X_{i} \not= a$
for any $i \in \mathcal{I} \setminus \Lambda$.
Hence
\begin{equation}
    \probability{A_{\Lambda}}
    = ( p ( a ) - p ( a ) w ( b | a ) )^{| \Lambda |}
    ( 1 - p ( a ) )^{n - | \Lambda |}
    \label{e:lemmaproofs.p0m.ppp} 
\end{equation}
for $\Lambda \subseteq \mathcal{I}$.
By the binomial formula we have
\begin{align}
    & \sum_{m = 1}^{n}
    \binom{n}{m}
    ( p ( a ) - p ( a ) w ( b | a ) )^{m}
    ( 1 - p ( a ) )^{n - m} \notag \\
    & = ( 1 - p ( a ) w ( b | a ) )^{n} - ( 1 - p ( a ) )^{n} .
    \label{e:lemmaproofs.p0m.binomial}
\end{align}
The inequality \eqref{e:sketch.p0m} follows from \eqref{e:lemmaproofs.p0m.psp},
\eqref{e:lemmaproofs.p0m.ppp} and \eqref{e:lemmaproofs.p0m.binomial}.

\subsection{Proof of Lemma~\ref{lem:sketch.chlmr}}
\label{u:lemmaproofs.chlmr}

For every $\theta \in \oointerval{0}{\infty}$ and function
$a : \mathcal{X} \to \reals$ we have \cite{merhav1994, scarlett2020}
\begin{align}
    & \lmrate ( p , w , k ) \notag \\
    & \ge \meanop{( X , Y ) \sim p \times w} \log_{2} \left( \frac{
        k^{\theta} ( X , Y ) 2^{a ( X )}
    }{
        \sum_{x \in \mathcal{X}} p ( x ) k^{\theta} ( x , Y ) 2^{a ( x )}
    } \right) .
\end{align}
Letting $\bar{\mathcal{X}} = \{ x \in \mathcal{X} | p ( x ) > 0 \}$,
$\theta = 1$ and $a ( x ) = - \log_{2} ( p ( x ) )$ for
$x \in \bar{\mathcal{X}}$, we get
\begin{equation}
    \lmrate ( p , w , k )
    \ge \meanop{( X , Y ) \sim p \times w} \log_{2} \left( \frac{
        k ( X , Y )
    }{
        p ( X ) \sum_{x \in \bar{\mathcal{X}}} k ( x , Y )
    } \right) .
\end{equation}
For $x \in \mathcal{X}$ and $y \in \mathcal{Y}$ satisfying
$p ( x ) w ( y | x ) > 0$, \eqref{e:sketch.chlmr} implies
\begin{equation}
    \log_{2} \left( \frac{
        k ( x , y )
    }{
        p ( x ) \sum_{x' \in \bar{\mathcal{X}}} k ( x' , y )
    } \right)
    > \log_{2} \left( \frac{w ( y | x )}{p w ( y )} \right)
    - \frac{\epsilon}{| \mathcal{Y} | p w ( y )} .
    \label{e:lemmaproofs.chlmr.logarithms}
\end{equation}
The expectation of \eqref{e:lemmaproofs.chlmr.logarithms} with $x$, $y$ and $x'$
replaced by $X$, $Y$ and $x$ respectively is
\begin{equation}
    \meanop{( X , Y ) \sim p \times w} \log_{2} \left(
        \frac{k ( X , Y )}{p ( X ) \sum_{x \in \bar{\mathcal{X}}} k ( x , Y )}
    \right)
    > \mimap ( p , w ) - \epsilon .
\end{equation}
This proves $\mimap ( p , w ) - \epsilon < \lmrate ( p , w , k )$.

\subsection{Proof of Lemma~\ref{lem:lemmaproofs.eeerror}}
\label{u:lemmaproofs.eeerror}

Define $
    \hat{H}
    = \hat{H}_{\mathrm{MM}} ( \{ Z_{i} \}_{i = 1}^{n} , | \mathcal{Z} | )
$.
By Proposition~1 in \cite{paninski2003},
\begin{align}
    - \log_{2} \left( 1 + \frac{| \mathcal{Z} | - 1}{n} \right)
    & \le \mean{\hat{H}}
    - \frac{| \mathcal{Z} | - 1}{2 n} \log_{2} ( e )
    - \entropy{Z_{1}} \notag \\
    & \le 0 . \label{e:lemmaproofs.eeerror.bias}
\end{align}
By the remark in Section~2 of \cite{antos2001},
\begin{equation}
    | \hat{H} - \mean{\hat{H}} |
    \le \log_{2} ( n ) \sqrt{\frac{2}{n} \ln \left( \frac{2}{\delta} \right)}
    \label{e:lemmaproofs.eeerror.deviation}
\end{equation}
with a probability no less than $1 - \delta$.
Since $\ln ( 1 + u ) \le u$ for all $u \in \oointerval{- 1}{\infty}$, we have
\begin{equation}
    \log_{2} \left( 1 + \frac{| \mathcal{Z} | - 1}{n} \right)
    \le \log_{2} ( e ) \frac{| \mathcal{Z} | - 1}{n} .
    \label{e:lemmaproofs.eeerror.logbound}
\end{equation}
Since \eqref{e:lemmaproofs.eeerror} holds whenever
\eqref{e:lemmaproofs.eeerror.bias}, \eqref{e:lemmaproofs.eeerror.deviation} and
\eqref{e:lemmaproofs.eeerror.logbound} hold, \eqref{e:lemmaproofs.eeerror}
holds with a probability no less than $1 - \delta$.

    \section{Proofs of the Theorems in Section~\ref{s:results}}
    \label{s:theoremproofs} 
    \subsection{Proof of Theorem~\ref{thm:results.pia}}
\label{u:theoremproofs.pia}

Let $a$ and $a'$ be two different elements of $\mathcal{X}$.
Let $b$ and $b'$ be two different elements of $\mathcal{Y}$.
Since $\epsilon < 1$ and $0 < \delta < 1 / 2$, there exists
$\tau \in \oointerval{0}{1}$ satisfying $1 - \bemap ( \tau ) > \epsilon$
and $\tau < 2 ( 1 - \sqrt[n]{\delta + 2^{- n}} )$, where $
    \bemap ( \tau )
    = - \tau \log_{2} ( \tau ) - ( 1 - \tau ) \log_{2} ( 1 - \tau )
$.
Define a PMF $p$ on $\mathcal{X}$ and a transition function $w$ from
$\mathcal{X}$ to $\mathcal{Y}$ by
\begin{equation}
    p ( x )
    = \begin{cases}
        1 / 2 , & x \in \{ a , a' \} , \\
        0 , & x \not\in \{ a , a' \}
    \end{cases}
\end{equation}
and
\begin{equation}
    w ( y | x )
    = \begin{cases}
        \tau , & x = a , y = b , \\
        1 - \tau , & x = a , y = b' , \\
        1 - \tau , & x \not= a , y = b , \\
        \tau , & x \not= a , y = b' , \\
        0 , & y \not\in \{ b , b' \}
    \end{cases}
\end{equation}
respectively.
We can verify $\mimap ( p , w ) = 1 - \bemap ( \tau ) > \epsilon$ and
\begin{equation}
    ( 1 - p ( a ) w ( b | a ) )^{n} - ( 1 - p ( a ) )^{n} > \delta .
    \label{e:theoremproofs.pia.powers}
\end{equation}
If $S$ is a sequence of $n$ i.i.d. pairs generated by $p \times w$ and a
plug-in algorithm returns $K$ for the input $S$, then \eqref{e:results.p0lmr}
holds because of Lemma~\ref{lem:sketch.c0lmr}, Lemma~\ref{lem:sketch.p0m} and
\eqref{e:theoremproofs.pia.powers}.


\subsection{Proof of Theorem~\ref{thm:results.vsa}}
\label{u:theoremproofs.vsa}

Let $S = \{ ( X_{i} , Y_{i} ) \}_{i = 1}^{n}$.
Define
\begin{equation}
    \mathcal{Z}
    = \{
        ( x , y ) \in \mathcal{X} \times \mathcal{Y}
    |
        p ( x ) w ( y | x ) > 0
    \} .
\end{equation}
For $( x , y ) \in \mathcal{Z}$, let $A_{x , y}$ denote the event
\begin{equation}
    2^{\epsilon / | \mathcal{Y} | p w ( y )} p w ( y ) K ( x , y )
    \le p ( x ) w ( y | x ) \sum_{x' \in \mathcal{X}} K ( x' , y ) .
    \label{e:theoremproofs.vsa.cllmr} 
\end{equation}
If $\mimap ( p , w ) - \epsilon \ge \lmrate ( p , w , K )$, then
Lemma~\ref{lem:sketch.chlmr} implies that \eqref{e:theoremproofs.vsa.cllmr}
holds for some $( x , y ) \in \mathcal{Z}$.
Hence
\begin{equation}
    1 - \probability{\mimap ( p , w ) - \epsilon < \lmrate ( p , w , K )}
    \le \sum_{( x , y ) \in \mathcal{Z}} \probability{A_{x , y}} .
    \label{e:theoremproofs.vsa.unionbound}
\end{equation}

We now fix $( x , y ) \in \mathcal{Z}$ and prove
\begin{equation}
    \probability{A_{x , y}} < \frac{\delta}{| \mathcal{X} | | \mathcal{Y} |} .
    \label{e:theoremproofs.vsa.pfxy} 
\end{equation}
Define $\zeta = p ( x ) w ( y | x )$ and
$\eta = 2^{\epsilon / | \mathcal{Y} | p w ( y )} p w ( y )$.
For every positive integer $i \le n$, define a random variable
\begin{equation}
    U_{i}
    = \begin{cases}
        0 , & Y_{i} \not= y , \\
        \zeta , & X_{i} \not= x , Y_{i} = y , \\
        \zeta - \eta , & X_{i} = x , Y_{i} = y .
    \end{cases}
\end{equation}
Then
\begin{equation}
    \mean{U_{i}}
    = \zeta p w ( y ) ( 1 - 2^{\epsilon / | \mathcal{Y} | p w ( y )} )
    \label{e:theoremproofs.vsa.meanu}
\end{equation}
for all positive integers $i \le n$ and
\begin{equation}
    \sum_{i = 1}^{n} U_{i}
    ={} \zeta \sum_{x' \in \mathcal{X}} \ocnumber{x' , y}{S}
    - \eta \ocnumber{x , y}{S}
\end{equation}
with probability 1.
Also note \eqref{e:theoremproofs.vsa.cllmr} is equivalent to
\begin{equation}
    \eta ( \ocnumber{x , y}{S} + n^{\alpha} )
    \le \zeta
    \sum_{x' \in \mathcal{X}}
    ( \ocnumber{x' , y}{S} + n^{\alpha} ) .
\end{equation}
Hence $A_{x , y}$ equals the event
$\sum_{i = 1}^{n} U_{i} \ge n^{\alpha} ( \eta - \zeta | \mathcal{X} | )$.
By \eqref{e:results.scac}, \eqref{e:theoremproofs.vsa.meanu} and
\begin{equation}
    2^{\epsilon / | \mathcal{Y} | p w ( y )} - 1
    > \frac{\epsilon \ln ( 2 )}{| \mathcal{Y} | p w ( y )} ,
\end{equation}
we have
\begin{equation}
    n^{\alpha} ( \eta - \zeta | \mathcal{X} | ) - \sum_{i = 1}^{n} \mean{U_{i}}
    > n^{\alpha} \eta . \label{e:theoremproofs.vsa.meansum}
\end{equation}
Since $\zeta \le p w ( y ) < \eta$, $\zeta - \eta \le U_{i} \le \zeta$ with
probability 1 for all positive integers $i \le n$.
The inequalities \eqref{e:results.scpc}, \eqref{e:theoremproofs.vsa.meansum}
and the Hoeffding inequality \cite{hoeffding1963}
\begin{equation}
    \probability{A_{x , y}}
    \le \exp \left( - \frac{2}{n \eta^{2}} \left(
        n^{\alpha} ( \eta - \zeta | \mathcal{X} | )
        - \sum_{i = 1}^{n} \mean{U_{i}}
    \right)^{2} \right)
\end{equation}
lead to \eqref{e:theoremproofs.vsa.pfxy}.

The inequality \eqref{e:results.phlmr} follows from
\eqref{e:theoremproofs.vsa.unionbound}, \eqref{e:theoremproofs.vsa.pfxy} and
$| \mathcal{Z} | \le | \mathcal{X} | | \mathcal{Y} |$.


\subsection{Proof of Theorem~\ref{thm:results.vseea}}
\label{u:theoremproofs.vseea}

Define
$\rho : \integers_{> 0} \times \oointerval{0}{1} \to \oointerval{0}{\infty}$ by
\begin{align}
    \rho ( n , \delta )
    ={} & \frac{
        | \mathcal{X} | | \mathcal{Y} |
        + | \mathcal{X} |
        + | \mathcal{Y} |
        - 3
    }{2 n} \log_{2} ( e ) \notag \\
    & + 3
    \log_{2} ( n )
    \sqrt{\frac{2}{n} \ln \left( \frac{8}{\delta} \right)} .
\end{align}
There exists a function $
    \nu :
    \oointerval{1 / 2}{\infty}
    \times \oointerval{0}{\infty}
    \times \oointerval{0}{\infty}
    \times \oointerval{0}{1}
    \to \integers_{\ge 0}
$ such that for all $\alpha \in \oointerval{1 / 2}{\infty}$,
$\beta \in \oointerval{0}{\infty}$, $\epsilon \in \oointerval{0}{\infty}$,
$\delta \in \oointerval{0}{1}$ and $n \in \integers$ satisfying
$\alpha + \beta < 1$ and $n \ge \nu ( \alpha , \beta , \epsilon , \delta )$,
the inequalities
\begin{gather}
    \ln ( 2 ) n^{1 - \alpha} \left( n^{- \beta} - \rho ( n , \delta ) \right)
    \ge | \mathcal{X} | | \mathcal{Y} | , \\
    n^{- \beta} + \rho ( n , \delta )
    \le \epsilon \label{e:theoremproofs.vseea.scac}
\end{gather}
and
\begin{equation}
    \left(
        \frac{1}{2}
        \ln \left( \frac{4 | \mathcal{X} | | \mathcal{Y} |}{\delta} \right)
    \right)^{1 / ( 2 \alpha - 1 )}
    \le n \label{e:theoremproofs.vseea.scpc}
\end{equation}
hold.

Let $\alpha > 1 / 2$, $\beta > 0$, $\alpha + \beta < 1$, $\epsilon > 0$,
$0 < \delta < 1$, $p$ be a PMF on $\mathcal{X}$, $w$ be a transition function
from $\mathcal{X}$ to $\mathcal{Y}$, $S = \{ ( X_{i} , Y_{i} ) \}_{i = 1}^{n}$
be a sequence of i.i.d. random pairs generated by $p \times w$ and
$n \ge \nu ( \alpha , \beta , \epsilon , \delta )$.
Suppose the $( \alpha , \beta )$-VSEE algorithm returns $( K , R )$ for the
input $S$.
Define $
    \hat{H}_{X}
    = \hat{H}_{\mathrm{MM}} ( \{ X_{i} \}_{i = 1}^{n} , | \mathcal{X} | )
$, $
    \hat{H}_{Y}
    = \hat{H}_{\mathrm{MM}} ( \{ Y_{i} \}_{i = 1}^{n} , | \mathcal{Y} | )
$, $
    \hat{H}_{X , Y}
    = \hat{H}_{\mathrm{MM}} ( S , | \mathcal{X} | | \mathcal{Y} | )
$ and $\hat{I} = \hat{H}_{X} + \hat{H}_{Y} - \hat{H}_{X , Y}$.
By Lemma~\ref{lem:lemmaproofs.eeerror}, each of
\begin{gather}
    | \hat{H}_{X} - \entropymap ( p ) |
    \le \frac{| \mathcal{X} | - 1}{2 n} \log_{2} ( e )
    + \log_{2} ( n ) \sqrt{\frac{2}{n} \ln \left( \frac{8}{\delta} \right)} ,
    \label{e:theoremproofs.vseea.xentropy} \\
    | \hat{H}_{Y} - \entropymap ( p w ) |
    \le \frac{| \mathcal{Y} | - 1}{2 n} \log_{2} ( e )
    + \log_{2} ( n ) \sqrt{\frac{2}{n} \ln \left( \frac{8}{\delta} \right)}
    \label{e:theoremproofs.vseea.yentropy}
\end{gather}
and
\begin{align}
    | \hat{H}_{X , Y} - \entropymap ( p \times w ) |
    \le{} & \frac{| \mathcal{X} | | \mathcal{Y} | - 1}{2 n} \log_{2} ( e )
    \notag \\
    & + \log_{2} ( n ) \sqrt{\frac{2}{n} \ln \left( \frac{8}{\delta} \right)}
    \label{e:theoremproofs.vseea.jointentropy}
\end{align}
holds with a probability no less than $1 - \delta / 4$.
Since the $\alpha$-virtual sample algorithm returns $K$ for the input $S$,
\begin{equation}
    n
    \ge \left( \frac{
        | \mathcal{X} | | \mathcal{Y} |
    }{
        \ln ( 2 ) ( n^{- \beta} - \rho ( n , \delta ) )
    } \right)^{1 / ( 1 - \alpha )}
\end{equation}
and \eqref{e:theoremproofs.vseea.scpc} holds, the probability of
\begin{equation}
    \mimap ( p , w ) - n^{- \beta} + \rho ( n , \delta )
    < \lmrate ( p , w , K )
    \label{e:theoremproofs.vseea.pvsa}
\end{equation}
is greater than $1 - \delta / 4$.
The inequalities \eqref{e:theoremproofs.vseea.xentropy},
\eqref{e:theoremproofs.vseea.yentropy} and
\eqref{e:theoremproofs.vseea.jointentropy} imply
$| \hat{I} - \mimap ( p , w ) | \le \rho ( n , \delta )$, which together with
\eqref{e:theoremproofs.vseea.scac}, \eqref{e:theoremproofs.vseea.pvsa} and
$R = \hat{I} - n^{- \beta}$ implies
\begin{equation}
    \mimap ( p , w ) - \epsilon \le R \le \lmrate ( p , w , K ) .
    \label{e:theoremproofs.vseea.success}
\end{equation}
Therefore, \eqref{e:theoremproofs.vseea.success} holds with a probability
greater than $1 - \delta$.

    \section{%
    Techniques for the Evaluations of the Plug-In Algorithm and Virtual Sample
    Algorithm}
\label{s:techniques}

In Section~\ref{u:evaluations.vsa}, we showed the CDFs of
$\lmrate ( p , w , K_{\mathrm{PI}} )$ and
$\lmrate ( p , w , K_{\mathrm{VS}} )$, the LM rates of the decoding metrics
returned by the plug-in algorithm and the virtual sample algorithm.
The computation of the CDFs is possible because the LM rates are determined by
$\{ \ocnumber{x , y}{S} \}_{x \in \mathcal{X} , y \in \mathcal{Y}}$, which
takes values on a finite set.
Define a sequence
$\{ ( a_{j} , b_{j} ) \}_{j = 0}^{| \mathcal{X} | | \mathcal{Y} | - 1}$ such
that
\begin{equation}
    \mathcal{X} \times \mathcal{Y}
    = \{
        ( a_{j} , b_{j} )
    |
        j \in \integers_{\ge 0} , j < | \mathcal{X} | | \mathcal{Y} |
    \} . \label{e:techniques.alphabet}
\end{equation}
Let $C$ be the set of all $
    ( c_{0} , c_{1} , \cdots , c_{| \mathcal{X} | | \mathcal{Y} | - 1} )
    \in \integers_{\ge 0}^{| \mathcal{X} | | \mathcal{Y} |}
$ satisfying
\begin{equation}
    \sum_{j = 0}^{| \mathcal{X} | | \mathcal{Y} | - 1} c_{j} = n .
    \label{e:techniques.sum}
\end{equation}
For every $\{ c_{j} \}_{j = 0}^{| \mathcal{X} | | \mathcal{Y} | - 1} \in C$,
\begin{equation}
    \{
        \ocnumber{a_{j} , b_{j}}{S}
    \}_{j = 0}^{| \mathcal{X} | | \mathcal{Y} | - 1}
    = \{ c_{j} \}_{j = 0}^{| \mathcal{X} | | \mathcal{Y} | - 1}
    \label{e:techniques.occnumbers}
\end{equation}
holds with the probability
\begin{equation}
    \frac{n !}{\prod_{j = 0}^{| \mathcal{X} | | \mathcal{Y} | - 1} c_{j} !}
    \prod_{j = 0}^{| \mathcal{X} | | \mathcal{Y} | - 1}
    p^{c_{j}} ( a_{j} )
    w^{c_{j}} ( b_{j} | a_{j} )
\end{equation}
and we can compute $\lmrate ( p , w , K_{\mathrm{PI}} )$ and
$\lmrate ( p , w , K_{\mathrm{VS}} )$ for the training sets $S$ satisfying
\eqref{e:techniques.occnumbers}.
This yields the CDFs of $\lmrate ( p , w , K_{\mathrm{PI}} )$ and
$\lmrate ( p , w , K_{\mathrm{VS}} )$.

\begin{algorithm}[!t]
    \KwIn{%
        Learning algorithm $A$, finite sets $\mathcal{X}$ and $\mathcal{Y}$,
        PMF $p$ on $\mathcal{X}$, transition function $w$ from $\mathcal{X}$ to
        $\mathcal{Y}$, positive integer $n$.%
    }
    \KwOut{PMF $q$ on $\cointerval{0}{\infty}$.}
    $q \gets$ the function mapping every non-negative number to 0\;
    Find a sequence
    $\{ ( a_{j} , b_{j} ) \}_{j = 0}^{| \mathcal{X} | | \mathcal{Y} | - 1}$
    satisfying \eqref{e:techniques.alphabet}\;
    $c_{0} \gets n$\;
    \For{$j \in \{ 1 , 2 , \cdots , | \mathcal{X} | | \mathcal{Y} | - 1 \}$}{
        $c_{j} \gets 0$\;
    }
    \While{$c_{| \mathcal{X} | | \mathcal{Y} | - 1} < n$}{
        $s \gets$ the empty sequence\;
        $u \gets n!$\;
        \For{%
            $j \in \{ 0 , 1 , \cdots , | \mathcal{X} | | \mathcal{Y} | - 1 \}$%
        }{
            Append $c_{j}$ $( a_{j} , b_{j} )$'s to $s$\;
            $
                u \gets
                u p^{c_{j}} ( a_{j} ) w^{c_{j}} ( b_{j} | a_{j} ) / ( c_{j} ! )
            $\;
        }
        $k \gets A ( s )$\;
        $q ( \lmrate ( p , w , k ) ) \gets q ( \lmrate ( p , w , k ) ) + u$\;
        $l \gets$ the least integer satisfying
        $0 \le l < | \mathcal{X} | | \mathcal{Y} |$ and $c_{l} > 0$\;
        $c \gets c_{l}$\;
        $c_{l} \gets 0$\;
        $c_{l + 1} \gets c_{l + 1} + 1$\;
        $c_{0} \gets c - 1$\;
    }
    $s \gets$ the sequence of $n$ $(
        a_{| \mathcal{X} | | \mathcal{Y} | - 1} ,
        b_{| \mathcal{X} | | \mathcal{Y} | - 1}
    )$'s\;
    $k \gets A ( s )$\;
    $
        u
        \gets p^{n} ( a_{| \mathcal{X} | | \mathcal{Y} | - 1} ) w^{n} (
            b_{| \mathcal{X} | | \mathcal{Y} | - 1}
        |
            a_{| \mathcal{X} | | \mathcal{Y} | - 1}
        )
    $\;
    $q ( \lmrate ( p , w , k ) ) \gets q ( \lmrate ( p , w , k ) ) + u$\;
    \caption{Computing the PMF of the LM rate.}
    \label{alg:techniques.computing}
\end{algorithm}

To enumerate the elements of $C$, note there is a one-to-one correspondence
from $C$ to the set $V$ of all
$\sum_{j = 0}^{| \mathcal{X} | | \mathcal{Y} | - 1} c_{j} ( n + 1 )^{j}$, where
$c_{0}$, $c_{1}$, $\cdots$,
$c_{| \mathcal{X} | | \mathcal{Y} | - 1} \in \integers_{\ge 0}$ satisfy
\eqref{e:techniques.sum}.
The following lemma shows how we can enumerate the elements of $V$.

\begin{lemma}
    \label{lem:techniques.enumeration}
    Let $J$, $n \in \integers_{> 0}$ and $V$ be the set of all
    \begin{equation}
        v = \sum_{j = 0}^{J - 1} c_{j} ( n + 1 )^{j} ,
        \label{e:techniques.number}
    \end{equation}
    where $c_{0}$, $c_{1}$, $\cdots$, $c_{J - 1} \in \integers_{\ge 0}$ satisfy
    \begin{equation}
        \sum_{j = 0}^{J - 1} c_{j} = n . \label{e:techniques.generalsum}
    \end{equation}
    If $c_{0}$, $c_{1}$, $\cdots$, $c_{J - 1} \in \integers_{\ge 0}$ satisfy
    \eqref{e:techniques.generalsum} and the number $v$ defined by
    \eqref{e:techniques.number} is not the greatest number in $V$, then the
    least number greater than $v$ in $V$ is
    \begin{align}
        v'
        ={} & c_{l}
        - 1
        + ( c_{l + 1} + 1 ) ( n + 1 )^{l + 1} \notag \\
        & + \sum_{j \in \integers , l + 2 \le j < J} c_{j} ( n + 1 )^{j} ,
    \end{align}
    where $l$ is the least integer satisfying $0 \le l < J$ and $c_{l} > 0$.
\end{lemma}

\begin{IEEEproof}
    By definition, $v'$ is in $V$.
    Since $v$ is the $( n + 1 )$-ary number
    $c_{J - 1} \cdots c_{l + 2} c_{l + 1} c_{l} \cdots c_{1} c_{0}$ and $v'$ is
    the $( n + 1 )$-ary number
    $c_{J - 1} \cdots c_{l + 2} ( c_{l + 1} + 1 ) 0 \cdots 0 ( c_{l} - 1 )$,
    $v' > v$.
    In the following we assume $v'' \in V$ and $v'' > v$ and prove
    $v'' \ge v'$.

    There exist $c_{0}''$, $c_{1}''$, $\cdots$,
    $c_{J - 1}'' \in \integers_{\ge 0}$ satisfying
    $\sum_{j = 0}^{J - 1} c_{j}'' = n$ and
    \begin{equation}
        v'' = \sum_{j = 0}^{J - 1} c_{j}'' ( n + 1 )^{j} .
    \end{equation}
    Let $m$ be the greatest integer satisfying $0 \le m < J$ and
    $c_{m}'' > c_{m}$.
    For $j \in \integers$ satisfying $m < j < J$, because $c_{j}'' > c_{j}$ is
    impossible and $c_{j}'' < c_{j}$ contradicts the assumption $v'' > v$, we
    have $c_{j}'' = c_{j}$.
    This observation, $c_{m}'' > c_{m}$ and
    $\sum_{j = 0}^{J - 1} c_{j}'' = \sum_{j = 0}^{J - 1} c_{j}$ lead to
    $\sum_{j = 0}^{m - 1} c_{j} > 0$, which implies $m > l$.
    The fact that $c_{j}'' = c_{j}$ for all $j \in \integers$ satisfying
    $m < j < J$ also leads to
    \begin{equation}
        \sum_{j = 0}^{m} c_{j}'' = \sum_{j = 0}^{m} c_{j} .
        \label{e:techniques.equalsums}
    \end{equation}
    Since $( n + 1 )^{j} \ge 1$ for all $j \in \integers_{\ge 0}$,
    \begin{equation}
        v''
        \ge \sum_{j = 0}^{m - 1} c_{j}''
        + \sum_{j = m}^{J - 1} c_{j}'' ( n + 1 )^{j} .
        \label{e:techniques.numbersums} 
    \end{equation}
    If $m = l + 1$ and $c_{m}'' = c_{m} + 1$, then we have
    $\sum_{j = 0}^{m} c_{j} = c_{l} + c_{m}'' - 1$ and
    \begin{equation}
        \sum_{j = m}^{J - 1} c_{j}'' ( n + 1 )^{j}
        = ( c_{m} + 1 ) ( n + 1 )^{m}
        + \sum_{j \in \integers , l + 2 \le j < J} c_{j} ( n + 1 )^{j} ,
    \end{equation}
    which together with \eqref{e:techniques.equalsums} and
    \eqref{e:techniques.numbersums} imply
    \begin{equation}
        v''
        \ge c_{l}
        - 1
        + ( c_{m} + 1 ) ( n + 1 )^{m}
        + \sum_{j \in \integers , l + 2 \le j < J} c_{j} ( n + 1 )^{j} ,
    \end{equation}
    i.e. $v'' \ge v'$.
    If $m > l + 1$ or $c_{m}'' > c_{m} + 1$, then we can see $v'' > v'$ by
    comparing the $( n + 1 )$-ary forms of $v''$ and $v'$.
\end{IEEEproof}

If we input the plug-in algorithm, $\mathcal{X}$, $\mathcal{Y}$, $p$, $w$ and
$n$ to Algorithm~\ref{alg:techniques.computing}, then the PMF returned by the
algorithm is a PMF of $\lmrate ( p , w , K_{\mathrm{PI}} )$.
The same is true for the virtual sample algorithm.

\end{document}